\documentclass[aps,reprint,superscriptaddress,preprintnumbers,nobibnotes,nofootinbib,amsmath,amssymb,longbibliography]{revtex4-2}

\usepackage[T1]{fontenc}
\usepackage[colorlinks=true,citecolor=blue,linkcolor=blue]{hyperref}
\usepackage[capitalise]{cleveref}
\usepackage{booktabs}
\usepackage{dsfont}
\usepackage{graphicx}
\usepackage[retain-zero-uncertainty=true, range-phrase=--, range-units=single]{siunitx}
\usepackage{pifont}
\usepackage[dvipsnames]{xcolor}
\usepackage[normalem]{ulem}

\begin{document}

\title{The ultra-high-energy event KM3-230213A within the global neutrino landscape}

\begin{abstract}
On February 13th, 2023, the KM3NeT/ARCA telescope detected a neutrino candidate with an estimated energy in the hundreds of PeVs. In this article, the observation of this ultra-high-energy neutrino is discussed in light of null observations above tens of PeV from the IceCube and Pierre Auger observatories. Performing a joint fit of all experiments under the assumption of an isotropic $E^{-2}$ flux, the best-fit single-flavour flux normalisation is $E^2 \Phi^{\rm 1f}_{\nu + \bar \nu} = 7.5 \times 10^{-10}~{\rm GeV cm^{-2} s^{-1} sr^{-1}}$ in the 90\% energy range of the KM3NeT event.  Furthermore, the ultra-high-energy data are then fit together with the IceCube measurements at lower energies, either with a single power law or with a broken power law, allowing for the presence of a new component in the spectrum. 
The joint fit including non-observations by other experiments in the ultra-high-energy region shows a slight preference for a break in the PeV regime if the ``High-Energy Starting Events'' sample is included, and no such preference for the other two IceCube samples investigated.
A stronger preference for a break appears if only the KM3NeT data is considered in the ultra-high-energy region, though the flux resulting from such a fit would be inconsistent with null observations from IceCube and Pierre Auger.
In all cases, the observed tension between KM3NeT and other datasets is of the order of $2.5\sigma-3\sigma$, and increased statistics are required to resolve this apparent tension and better characterise the neutrino landscape at ultra-high energies.
\end{abstract}

\collaboration{The KM3NeT Collaboration}

\author{O.~Adriani}
\affiliation{INFN, Sezione di Firenze, via Sansone 1, Sesto Fiorentino, 50019 Italy}
\affiliation{Universit{\`a} di Firenze, Dipartimento di Fisica e Astronomia, via Sansone 1, Sesto Fiorentino, 50019 Italy}
\author{S.~Aiello}
\affiliation{INFN, Sezione di Catania, (INFN-CT) Via Santa Sofia 64, Catania, 95123 Italy}
\author{A.~Albert}
\affiliation{Universit{\'e}~de~Strasbourg,~CNRS,~IPHC~UMR~7178,~F-67000~Strasbourg,~France}
\affiliation{Universit{\'e} de Haute Alsace, rue des Fr{\`e}res Lumi{\`e}re, 68093 Mulhouse Cedex, France}
\author{A.\,R.~Alhebsi}
\affiliation{Khalifa University of Science and Technology, Department of Physics, PO Box 127788, Abu Dhabi,   United Arab Emirates}
\author{M.~Alshamsi}
\affiliation{Aix~Marseille~Univ,~CNRS/IN2P3,~CPPM,~Marseille,~France}
\author{S.~Alves~Garre}
\affiliation{IFIC - Instituto de F{\'\i}sica Corpuscular (CSIC - Universitat de Val{\`e}ncia), c/Catedr{\'a}tico Jos{\'e} Beltr{\'a}n, 2, 46980 Paterna, Valencia, Spain}
\author{A.~Ambrosone}
\affiliation{Universit{\`a} di Napoli ``Federico II'', Dip. Scienze Fisiche ``E. Pancini'', Complesso Universitario di Monte S. Angelo, Via Cintia ed. G, Napoli, 80126 Italy}
\affiliation{INFN, Sezione di Napoli, Complesso Universitario di Monte S. Angelo, Via Cintia ed. G, Napoli, 80126 Italy}
\author{F.~Ameli}
\affiliation{INFN, Sezione di Roma, Piazzale Aldo Moro 2, Roma, 00185 Italy}
\author{M.~Andre}
\affiliation{Universitat Polit{\`e}cnica de Catalunya, Laboratori d'Aplicacions Bioac{\'u}stiques, Centre Tecnol{\`o}gic de Vilanova i la Geltr{\'u}, Avda. Rambla Exposici{\'o}, s/n, Vilanova i la Geltr{\'u}, 08800 Spain}
\author{L.~Aphecetche}
\affiliation{Subatech, IMT Atlantique, IN2P3-CNRS, Nantes Universit{\'e}, 4 rue Alfred Kastler - La Chantrerie, Nantes, BP 20722 44307 France}
% ...
\date{February 2025}

\maketitle

\section{Introduction}
\label{sec:intro}

The highest-energy neutrino ever observed, KM3-230213A~\cite{Nature}, has been recently detected with a partial configuration of the KM3NeT/ARCA detector~\cite{KM3Net:2016zxf,KM3NeT:2022pnv}. The energy of the event is $E_\nu = \SI[parse-numbers=false]{220^{+570}_{-110}}{\peta\eV}$, i.e., falling in the ultra-high-energy (UHE) region ($\gtrsim~\SI{50}{\peta\eV}$), surpassing prior observations by the IceCube Neutrino Observatory (IC,~\cite{Meier:2024flg}) by one order of magnitude. Given the steeply falling atmospheric neutrino spectrum~\cite{Illana:2010gh,Ostapchenko:2022thy} and the arrival direction of the event, an astrophysical origin for KM3-230213A is most likely.

Though the origin of the event has not been identified so far~\cite{Nature}, its provenance may fall into one of the following broad categories: steady source, transient source, cosmogenic origin, or beyond the Standard Model. In this article, we only consider the scenario in which the event is associated with an isotropic ($4\pi$ solid angle) diffuse neutrino flux, regardless of its origin, and leave the study of other peculiar cases for separate works. Within this hypothesis, we combine the KM3NeT observation with searches for high- and ultra-high-energy neutrinos performed by IC and the Pierre Auger Observatory (Auger,~\cite{PierreAuger:2023pjg}).

The KM3NeT exposure associated with the bright track selection described in Ref.~\cite{Nature} and with the data-taking period with 19 and 21 lines out of the envisaged 230~\cite{KM3NeT:2024paj} is defined as:
\begin{equation}
    \mathcal{E}^{\rm KM3}(E) = 4\pi \times T^{\rm KM3} \times A_{\rm eff}^{\rm KM3}(E),
    \label{eq:exposure:km3net}
\end{equation}
where $T^{\rm KM3}$ and $A_{\rm eff}^{\rm KM3}(E)$ are respectively the livetime of 335 days and the sky-averaged effective area spanning energies from \SI{100}{\tera\eV} to \SI{100}{\exa\eV}. The event selection is associated to a negligible background ($\mathcal{O}(10^{-5})$/year) from atmospheric origin \cite{Nature}, that is therefore neglected in the following computations. The IceCube and Auger datasets are detailed in \cref{sec:datasets}. Measurements by other experiments~\cite{Anker:2019rzo,ANITA:2019wyx,ARA:2019uvt} are not considered in our analysis, as IC and Auger are the most sensitive in the relevant energy region. To understand the implications of KM3-230213A within the context of global data, we perform a joint analysis of the neutrino flux from TeV to EeV, for both single and broken power law flux hypotheses and estimate the preference between these two scenarios. The methods introduced in this study constitute an important benchmark for future multi-experiment investigations of ultra-high-energy neutrinos.

The article is organised as follows. In~\cref{sec:compatibility} we assess the compatibility between the KM3-230213A observation and IC/Auger non-observations by combining all datasets at the KM3-230213A energy range. In~\cref{sec:jointfit}, we additionally incorporate the IC high-energy (HE) measurements for energies below \SI{10}{\peta\eV} obtained using High-Energy Starting Events (HESE,~\cite{IceCube:2020wum}), Enhanced Starting Track Event Selection (ESTES,~\cite{IceCube:2024fxo}), or Northern-Sky Tracks (NST,~\cite{IceCube:2021uhz}). Finally, in~\cref{sec:km3netfit}, we assess the compatibility between the IC HE measurements and only the KM3-230213A observation, before summarising our findings in \cref{sec:conclusion}.

\begin{figure*}[hbtp]
    \centering
    \includegraphics[width=\linewidth]{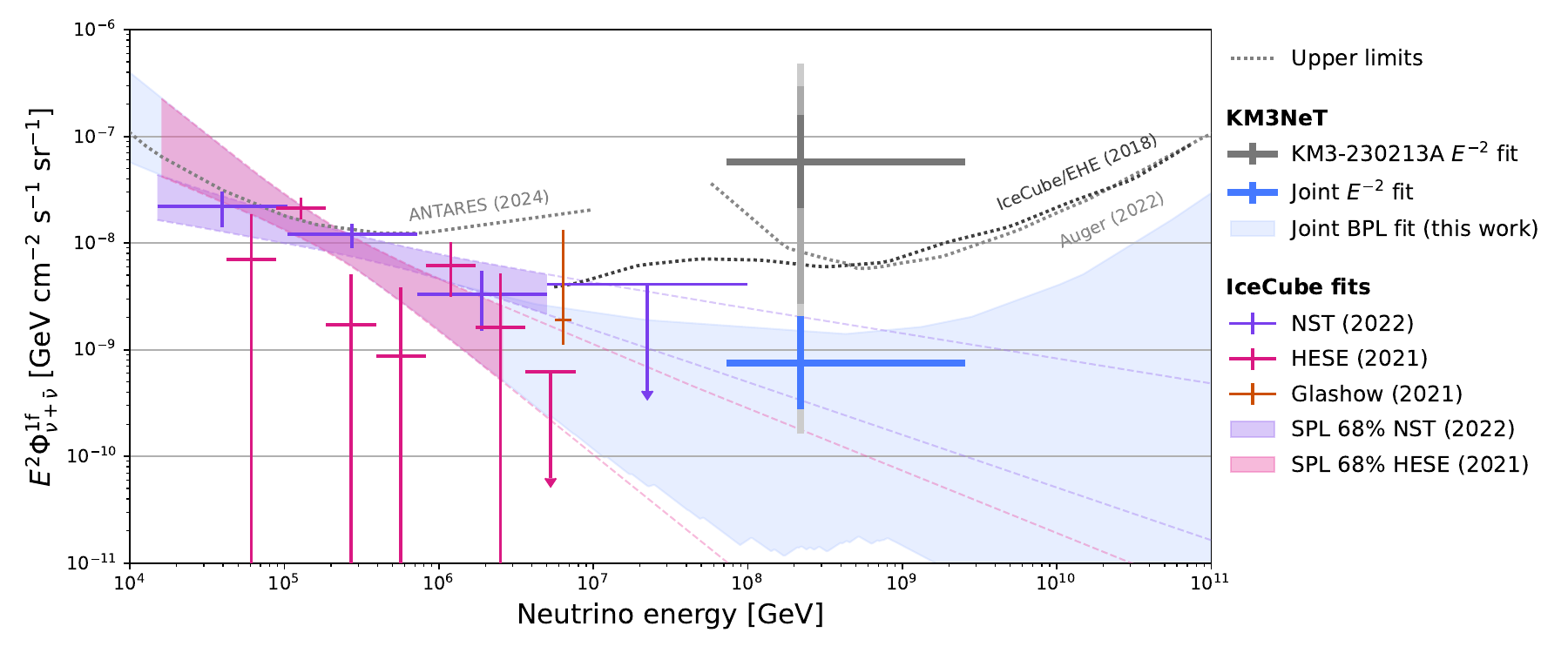}
    \caption{Per-flavour energy-squared diffuse astrophysical neutrino flux, assuming flavour equipartition. Measurements based on the KM3-230213A observation are compared with existing limits at ultra-high energies and measurements at lower energies. The KM3NeT-only measurement~\cite{Nature} and the joint flux including as well IC-EHE and Auger non-observations, in the central 90\% neutrino energy range associated with KM3-230213A, are shown with the grey and blue crosses, respectively. The purple- and pink-filled regions represent the 68\% CL contours of the IC single-power-law fits, NST~\cite{IceCube:2021uhz} and HESE~\cite{IceCube:2020wum}, respectively. The segmented fits from the same analyses are shown with purple and pink crosses, while the orange cross corresponds to the IC Glashow resonance event~\cite{IceCube:2021rpz}. The dotted lines correspond to upper limits from ANTARES (95\% CL~\cite{ANTARES:2024ihw}), Auger (90\% CL~\cite{PierreAuger:2023pjg}) and IC-EHE (90\% CL~\cite{IceCube:2018fhm}). The blue-filled region shows the $1\sigma$ envelope of the broken power law fit, using the HESE constraint below the break, as detailed in the text. The vertical spans of all crosses represent the $1\sigma$ uncertainties, and the $2-3\sigma$ ones for the KM3NeT-only point, in lighter shades.\label{fig:flux}}
\end{figure*}

\section{Compatibility with UHE landscape}
\label{sec:compatibility}

As detailed in Ref.~\cite{Nature}, the observation of KM3-230213A can be associated to a per-flavour isotropic diffuse flux measurement of $E^2 \Phi^{\rm 1f}_{\nu+\bar\nu}(E) = 5.8^{+10.1}_{-3.7}\times\SI{e-8}{\GeV\per\square\cm\per\second\per\steradian}$, when assuming an $E^{-2}$ spectrum in the central 90\% energy range of the event ($\SI{72}{\peta\eV} - \SI{2.6}{\exa\eV}$) and equipartition of neutrino flavours. If the non-observations in the IC Extremely-High-Energy sample (IC-EHE,~\cite{IceCube:2018fhm}) and Auger~\cite{PierreAuger:2023pjg}, detailed in \cref{sec:datasets}, are incorporated in the total exposure and following the same statistical methods as in Ref.~\cite{Nature}, the best-fit flux becomes $7.5^{+13.1}_{-4.7}\times\SI{e-10}{\GeV\per\square\cm\per\second\per\steradian}$. In \cref{fig:flux}, these values are compared with existing UHE constraints and IC HE observations. The probability of observing one event in KM3NeT and zero in both IC-EHE and Auger assuming the flux resulting from the joint fit is about $0.5\%$ ($2.6\sigma$). In the following, we extend our investigations to spectral shapes beyond an $E^{-2}$ spectrum.

\begin{figure*}
    \centering
    \includegraphics[width=\linewidth]{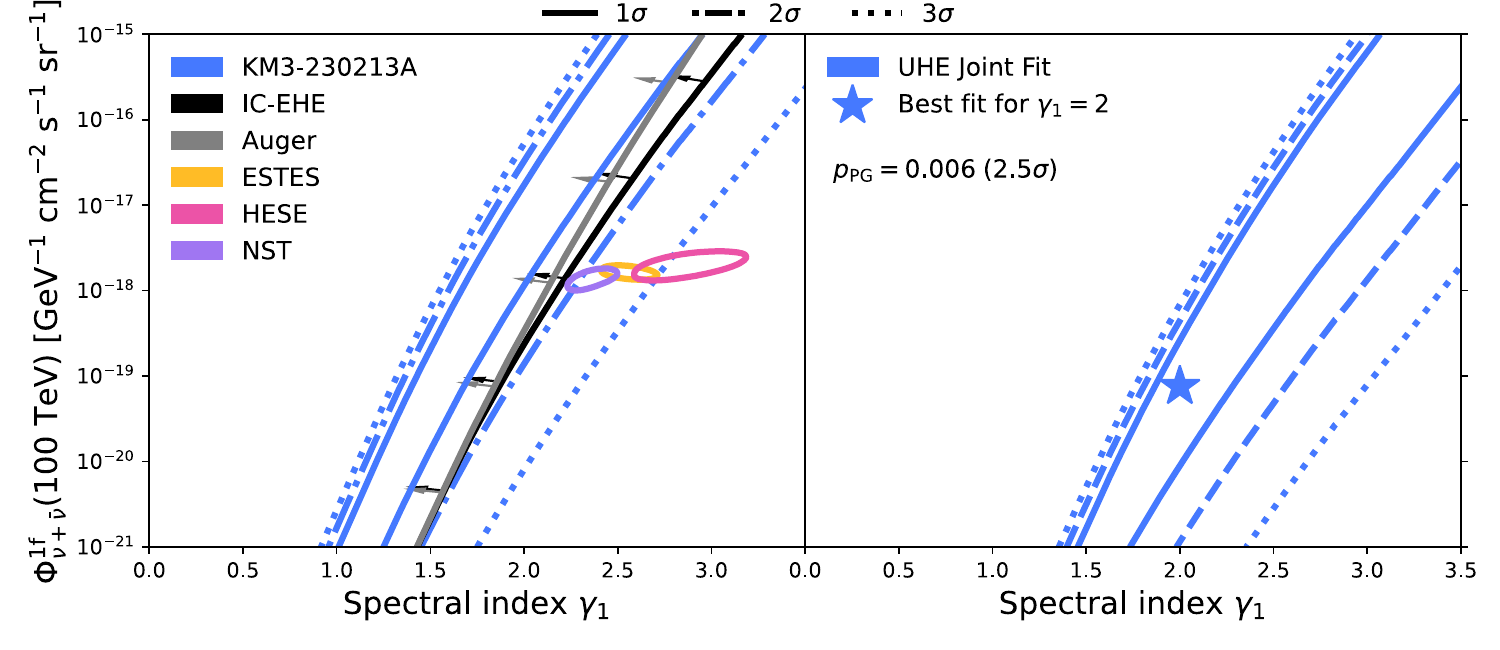}
    \caption{Fit results of the KM3NeT / IC / Auger data to a single-power-law flux model. Left: preferred regions and constraints from each dataset individually. The UHE datasets are all fit within the 90\% CL energy range of KM3-230213A, while the lower energy IC datasets are fit within their own defined energy ranges, always below that of the UHE datasets. The KM3-230213A contours are shown at the $1-2-3\sigma$ CLs (solid, dash-dotted and dotted lines, respectively), while all other contours are shown at the $1\sigma$ CL. Right: result of a joint fit to the UHE datasets. The star indicates the best-fit point for $\gamma_1=2$. The resulting tension from the PG test, detailed in the text, is also reported.\label{fig:spl-results}}
\end{figure*}

The compatibility with UHE non-observations by IC and Auger is estimated by performing a joint fit under the assumption of a single power law (SPL), parametrised as
\begin{equation}
\Phi^{\rm 1f}_{\nu+\bar\nu}(E; \Theta=\{\phi, \gamma_1\}) = \phi \times \left(\frac{E}{100\,{\rm TeV}}\right)^{-\gamma_1},
\end{equation}
where $\phi$ and $\gamma_1$ represent the flux normalisation and the spectral index, respectively. First, we use the likelihoods $\mathcal{L}^{\rm in}_d$, for each dataset $d$ (KM3NeT, IC-EHE, Auger) and within (``in'') the energy range of KM3-230213A, as defined in \cref{eq:lkl_in}, and Wilks' theorem~\cite{Wilks:1938dza} with two degrees of freedom (d.o.f.) to obtain contours in the SPL parameter space. The preferred region from KM3-230213A and constraints from IC-EHE and Auger are shown in the left panel of \cref{fig:spl-results}. Then we use the same approach with $\mathcal{L}^{\rm in}_{\rm UHE}(\Theta) = \prod_d \mathcal{L}^{\rm in}_d(\Theta)$ to perform the joint fit, as shown in the right panel of \cref{fig:spl-results}.

To quantify the tension between the UHE datasets, the parameter-goodness-of-fit (PG) test~\cite{Maltoni:2003pg} is performed. To compute this, we first define $\overline{\chi}^2_{\rm in}$ as:
\begin{equation}
\overline{\chi}^2_{\rm in}(\phi,\gamma_1) = \chi^2_{\rm in}(\phi,\gamma_1) - \sum_d \hat{\chi}^2_{d,\rm in}, \label{eq:chi2}
\end{equation}
with $\chi^2_{\rm in}(\phi,\gamma_1) = -2 \log \mathcal{L}^{\rm in}_{\rm UHE}(\phi,\gamma_1)$ and $\hat{\chi}^2_{d,\rm in} = \min_{(\phi,\gamma_1)} \left[ -2 \log{\mathcal{L}^{\rm in}_d(\phi,\gamma_1)} \right]$, where the sum runs over the three UHE datasets and the hatted parameters indicate the best-fit values, maximising the likelihood. The PG test statistic $t_{\rm PG} = \min_{(\phi,\gamma_1)} \overline{\chi}^2_{\rm in}(\phi,\gamma_1)$ follows a $\chi^2$ distribution with one d.o.f. We obtain a p-value $p_{\rm PG} = 0.006$ and a one-tailed z-score $z_{\rm PG} = 2.5\sigma$. This number stays relatively consistent with the $E^{2}$ value quoted at the beginning of the section and shows only a moderate tension between the UHE datasets.

Similarly, we define a Bayesian analysis probing the same compatibility. The posterior $P(\phi, \gamma_1)$ is the product of the UHE likelihood $\mathcal{L}^{\rm in}_{\rm UHE}(\phi, \gamma_1)$ and non-informative uniform priors $\mathcal{U}_{[0,\,\num{e-10}]}(\phi)$ (in \si{\per\GeV\per\square\cm\per\second\per\steradian}) and $\mathcal{U}_{[1, 4]}(\gamma_1)$, where $\mathcal{U}_{[a,\,b]}$ denotes the continuous uniform distribution between $a$ and $b$. The tension is computed using the posterior predictive check (PPC) approach \cite{Gelman:1996}. The joint probability is:
\begin{equation}
    p_{\rm PPC} = \iint P(\phi, \gamma_1) \times \prod_d \mathcal{L}_d^{\rm in}(\phi, \gamma_1) \, {\rm d}\phi {\rm d}\gamma_1,
    \label{eq:ppc}
\end{equation}
which can be converted to a z-score using the one-tailed convention. A final z-score of $2.5\sigma$ is obtained, similar to the value from the PG test, as reported in \cref{tab:tensions}.

The tension estimated in both approaches relies on minimal assumptions on the shape of the astrophysical flux, namely that such a flux is well modelled by an SPL within the energy range of interest. Therefore, this tension is a consequence of the difference in the current exposure at UHE between the three experiments.

\begin{table}[hbtp]
    \caption{Summary of the computed tension. Each row corresponds to one specific set of samples and the different columns correspond to different fit assumptions ($E^{-2}$ spectrum, SPL, BPL). In each cell, the values estimated with a parameter-goodness-of-fit test for the frequentist approach and a posterior predictive check of UHE observations for the Bayesian case, separated with a slash, are reported.\label{tab:tensions}}
    \begin{ruledtabular}
    \begin{tabular}{ccc|c|ccc}
    \multicolumn{4}{c|}{Samples\footnote{K: KM3NeT, I: IceCube, A: Auger, HE: IC measurements at high energies (H=HESE, E=ESTES, N=NST)}}  & \multicolumn{3}{c}{Flux assumption} \\
    \multicolumn{3}{c|}{UHE} & HE & & & \\
    K & I & A & I & $E^{-2}$ & SPL & BPL \\
    \midrule
    \ding{51} & \ding{51} & \ding{51} & \ding{53} & $2.7\sigma$ / $2.9\sigma$ & $2.5\sigma$ / $2.5\sigma$ & $-$ \\
    \midrule
    \ding{51} & \ding{51} & \ding{51} & H & $-$ & $2.0\sigma$ / $3.0\sigma$ & $2.9\sigma$ / $2.9\sigma$ \\
    \ding{51} & \ding{51} & \ding{51} & E & $-$ & $1.6\sigma$ / $2.8\sigma$ & $2.9\sigma$ / $2.9\sigma$ \\
    \ding{51} & \ding{51} & \ding{51} & N & $-$ & $1.6\sigma$ / $2.9\sigma$ & $2.9\sigma$ / $2.9\sigma$ \\
    \midrule
    \ding{51} & \ding{53} & \ding{53} & H & $-$ & $2.8\sigma$ / $2.7\sigma$ & $-$ \\
    \ding{51} & \ding{53} & \ding{53} & E & $-$ & $2.2\sigma$ / $2.3\sigma$ & $-$ \\
    \ding{51} & \ding{53} & \ding{53} & N & $-$ & $1.6\sigma$ / $1.9\sigma$ & $-$ \\
    \end{tabular}
    \end{ruledtabular}
\end{table}

\section{Joint global fit}
\label{sec:jointfit}

Measurements using the IC HE analyses ---HESE, ESTES, or NST--- can also be incorporated into the analysis so that the full energy range from TeV to EeV is considered. The UHE measurements are also extended to the full PeV--ZeV region, as detailed in \cref{sec:stat}. In this section we define $\mathcal{L}_{\rm UHE}(\Theta) = \prod_d \mathcal{L}_d(\Theta)$, where the product runs over KM3NeT, IC-EHE, and Auger, and $\mathcal{L}_d$ is defined in \cref{eq:lkl}.

\subsection{Single-power-law scenario}

First, the joint fit of HE and UHE measurements is done with the assumption that the flux follows an SPL spanning from TeV to EeV. The combined likelihood is
\begin{align}
    \mathcal{L}^{\rm(a)}_{\rm{SPL}}(\phi, \gamma_1) = \mathcal{L}_{\rm UHE}(\phi, \gamma_1) \times 
    \mathcal{L}^{\rm(a)}_{\rm IC}(\phi, \gamma_1),
    \label{eq:lkl_SPL}
\end{align}
where the second term is defined in \cref{eq:ts_ic}, and $a$ refers to any of the three IC HE samples.

The analysis is performed using a Bayesian approach, where the posterior $P^{(a)}(\phi, \gamma_1)$ is the product of the likelihood in \cref{eq:lkl_SPL} and the uniform priors, $\mathcal{U}_{[0,\,\num{4e-18}]}(\phi)$ (in \si{\per\GeV\per\square\cm\per\second\per\steradian}) and $\mathcal{U}_{[1, 4]}(\gamma_1)$. The procedure is repeated separately for the HESE, ESTES, and NST IC datasets; a combined fit including all these datasets is beyond our scope due to the non-trivial overlap between the datasets.

\begin{figure*}
    \centering
    \includegraphics[width=\linewidth]{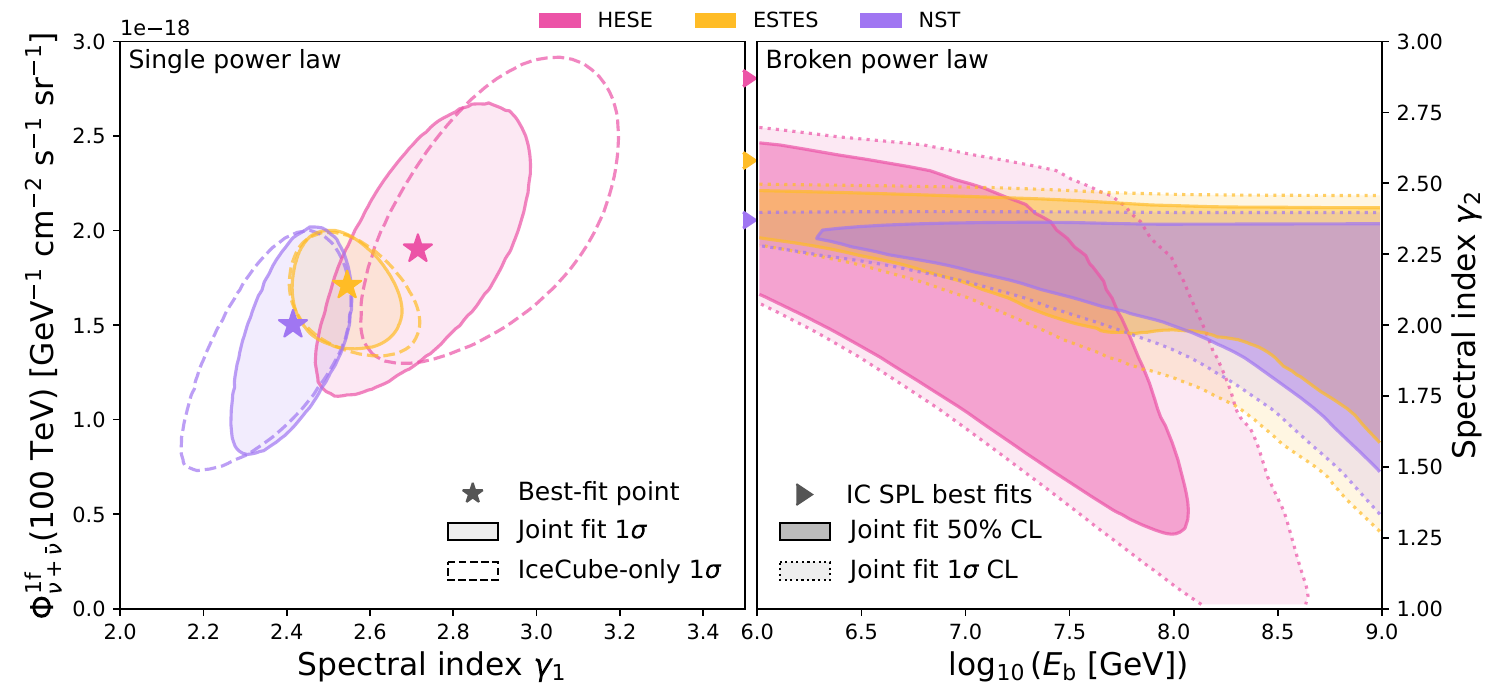}
    \caption{Joint fit of IC measurements in the TeV--PeV region, and KM3NeT / IC / Auger UHE measurements. Left: Bayesian fit with a single power law in the $(\gamma_1, \phi)$ plane. Right: broken power law fit in the $(\log_{10}E_b, \gamma_2)$ plane, after marginalising over $\phi$ and $\gamma_1$. The different colours correspond to different IC measurements: HESE in pink, ESTES in orange, and NST in purple. The filled regions indicate the $1\sigma$ contours (as well as the $50\%$ contours in the right panel). On the left panel, the star markers are the best-fit points and the $1\sigma$ contours from IC priors are also shown in dashed lines. On the right panel, the triangles on the left axis represent the value of the spectral indices corresponding to the IC SPL best fits.\label{fig:jointfit}}
\end{figure*}

The corresponding 2D contours are shown in the left panel of \cref{fig:jointfit}, compared to the IC measurements alone. Due to the large lever arm provided by the inclusion of UHE datasets, improved constraints on the slope of the SPL spectrum can be achieved. The best-fit spectral indices converge towards similar values, in line with the IC ``global fit''~\cite{Naab:2023xcz}. The best-fit flux are estimated by finding the maximum ($\hat\phi, \hat\gamma_1 = \textrm{arg max}_{(\phi,\gamma_1)}P^{(a)}(\phi, \gamma_1)$) and the Highest-Posterior-Density $1\sigma$ intervals are extracted from profiled posteriors. Alternatively, in the frequentist approach, we derive the 2D contours and 1D intervals using the likelihood defined in \cref{eq:lkl_SPL} and Wilks' theorem. The best-fit flux parameters obtained in both approaches are in good agreement, and only the results of the Bayesian fits are reported in \cref{tab:bestfits}.

Akin to the previous section, we compute the joint probability $p_{\rm ppc}$ of UHE observations as in \cref{eq:ppc} and we perform a PG test to quantify the compatibility between the different samples used for the SPL fit. For the latter, \cref{eq:chi2} is modified as
\begin{align}
    \overline{\chi}^2_{\rm (a)}(\phi,\gamma_1) = \overline{\chi}^2(\phi,\gamma_1) + \textrm{TS}^{\rm (a)}_{\rm IC}(\phi, \gamma_1) - \min\textrm{TS}^{\rm (a)}_{\rm IC},
    \label{eq:chi2mod}
\end{align}
where $\textrm{TS}^{\rm (a)}_{\rm IC}$ is defined in \cref{eq:ts_ic} and $\overline{\chi}^2 = -2\log\mathcal{L}_{\rm UHE} - \sum_d \min\left[-2 \log\mathcal{L}_d\right]$. The obtained results are summarised in \cref{tab:tensions}. The PPC numbers ($2.8\sigma-3.0\sigma$) are similar to the one from the previous section, reflecting the tension between the UHE datasets. Instead, the tension is relaxed with the PG test ($1.6\sigma-2.0\sigma$), as the IC-EHE and Auger non-observations are in line with the SPL flux extrapolated from IC HE measurements and only KM3-230213A is an outlier.

\begin{table*}[hbtp]
    \renewcommand{\arraystretch}{1.1}
    \caption{Summary of the main results of the joint fit of the UHE landscape. Each row corresponds to one specific fit. The sample(s) included in the fit is (are) indicated in the first four columns. The assumed flux hypothesis ($E^{-2}$, SPL, BPL) is given in the next column. The best-fit values and the corresponding $1\sigma$ intervals on the free parameters are presented in the last four columns. While the first two rows~\cite{Nature} are obtained in a frequentist approach, the other rows report the Bayesian results, that are also fully consistent with the maximum-likelihood fits.\label{tab:bestfits}}
    \begin{minipage}{0.70\linewidth}
    \begin{ruledtabular}
    \begin{tabular}{ccc|c|c|cccc}
    \multicolumn{4}{c|}{Samples\footnote{K: KM3NeT, I: IceCube, A: Auger, HE: IC measurements at high energies (H=HESE, E=ESTES, N=NST)}} & Assumed & \multicolumn{4}{c}{Best-fit flux} \\
    \multicolumn{3}{c|}{UHE} & HE & flux & {$\phi=\Phi^{\rm 1f}_{\nu+\bar\nu}(\SI{100}{\tera\eV})$} & {$\gamma_1$} & {$\gamma_2$} & {$\log_{10}E_{\rm b}/\si{GeV}$} \\
    K & I & A & I & & {\si{\per\GeV\per\square\cm\per\second\per\steradian}} & & & \\
    \midrule
    \ding{51} & \ding{53} & \ding{53} & \ding{53} & $E^{-2}$ & $5.8^{+10.1}_{-3.7} \times \num{e-18}$ & {$-$} & {$-$} & {$-$} \\
    \ding{51} & \ding{51} & \ding{51} & \ding{53} & $E^{-2}$ & $7.5^{+13.1}_{-4.7} \times \num{e-20}$ & {$-$} & {$-$} & {$-$} \\
    \midrule
    \multicolumn{9}{c}{\textit{\textbf{Joint global fit}}} \\
    \midrule
    \ding{51} & \ding{51} & \ding{51} & H & SPL & $1.9^{+0.5}_{-0.5} \times \num{e-18}$ & $2.71^{+0.17}_{-0.15}$ & {$-$} & {$-$} \\
    \ding{51} & \ding{51} & \ding{51} & E & SPL & $1.7^{+0.2}_{-0.2} \times \num{e-18}$ & $2.54^{+0.08}_{-0.08}$ & {$-$} & {$-$} \\
    \ding{51} & \ding{51} & \ding{51} & N & SPL & $1.5^{+0.4}_{-0.4} \times \num{e-18}$ & $2.42^{+0.09}_{-0.09}$ & {$-$} & {$-$} \\
    \midrule
    \ding{51} & \ding{51} & \ding{51} & H & BPL & $2.1^{+0.5}_{-0.5} \times \num{e-18}$ & $2.87^{+0.17}_{-0.20}$ & $2.22^{+0.33}_{-0.90}$ & $6.4^{+1.3}_{-0.4}$ \\
    \ding{51} & \ding{51} & \ding{51} & E & BPL & $1.7^{+0.2}_{-0.2} \times \num{e-18}$ & $2.57^{+0.09}_{-0.08}$ & $2.22^{+0.34}_{-0.82}$ & $7.3^{+0.6}_{-1.3}$ \\
    \ding{51} & \ding{51} & \ding{51} & N & BPL & $1.5^{+0.3}_{-0.4} \times \num{e-18}$ & $2.41^{+0.10}_{-0.09}$ & $2.41^{+0.05}_{-0.80}$ & $7.4^{+1.5}_{-1.4}$ \\
    \midrule
    \multicolumn{9}{c}{\textit{\textbf{Joint fit of IC HE and KM3-230213A}}} \\
    \midrule
    \ding{51} & \ding{53} & \ding{53} & H & SPL & $1.9^{+0.4}_{-0.6} \times \num{e-18}$ & $2.70^{+0.16}_{-0.21}$ & {$-$} & {$-$} \\
    \ding{51} & \ding{53} & \ding{53} & E & SPL & $1.7^{+0.2}_{-0.2} \times \num{e-18}$ & $2.51^{+0.08}_{-0.12}$ & {$-$} & {$-$} \\
    \ding{51} & \ding{53} & \ding{53} & N & SPL & $1.3^{+0.4}_{-0.5} \times \num{e-18}$ & $2.26^{+0.15}_{-0.17}$ & {$-$} & {$-$} \\
    \midrule
    \ding{51} & \ding{53} & \ding{53} & H & BPL & $2.1^{+0.4}_{-0.5} \times \num{e-18}$ & $2.85^{+0.19}_{-0.16}$ & $1.02^{+0.51}_{-0.01}$ & $7.1^{+0.2}_{-0.8}$ \\
    \ding{51} & \ding{53} & \ding{53} & E & BPL & $1.7^{+0.3}_{-0.3} \times \num{e-18}$ & $2.57^{+0.07}_{-0.11}$ & $1.02^{+0.53}_{-0.02}$ & $7.4^{+0.3}_{-1.0}$ \\
    \ding{51} & \ding{53} & \ding{53} & N & BPL & $1.4^{+0.4}_{-0.4} \times \num{e-18}$ & $2.37^{+0.11}_{-0.15}$ & $1.03^{+0.55}_{-0.03}$ & $7.7^{+0.4}_{-1.1}$ \\
    \end{tabular}
    \end{ruledtabular}
    \end{minipage}
\end{table*}

\subsection{Broken-power-law scenario}

The energy of KM3-230213A, significantly higher than the energy of all neutrinos detected so far, motivates the exploration of an astrophysical neutrino flux with additional components beyond an SPL. A broken power law (BPL) flux is therefore considered:
\begin{equation}
    \Phi^{\rm 1f}_{\nu+\bar\nu}(E) = \phi \times \begin{cases}
        \left(\dfrac{E}{\SI{100}{\tera\eV}}\right)^{-\gamma_1}; \; E \leq E_{\rm b}, \\
        \left(\dfrac{E}{E_{\rm b}}\right)^{-\gamma_2} \left(\dfrac{E_{\rm b}}{\SI{100}{\tera\eV}}\right)^{-\gamma_1}; \; E > E_{\rm b},
    \end{cases}
    \label{eq:bpl}
\end{equation}
introducing two new parameters, $\gamma_2$ and $E_{\rm b}$, characterising the spectral index above the break and the energy of the break, respectively.

Given the motivation for the presence of new components, we limit the exploration of the parameter space to $\gamma_2 \in [1, \gamma_1]$, i.e., where the spectrum hardens after the break. Additionally, as the BPL fit should not affect IC measurements below \SI{1}{\peta\eV}, assumed to follow an SPL in the following, and as no measurements are available in the EeV region, the parameter space in $E_{\rm b}$ is limited to the $\SI{1}{\peta\eV} - \SI{1}{\exa\eV}$ range. The combined likelihood is
\begin{align}
    \mathcal{L}^{\rm (a)}_{\rm BPL}(\phi, \gamma_1, \gamma_2, E_{\rm b}) &= \mathcal{L}_{\rm UHE}(\phi, \gamma_1, \gamma_2, E_{\rm b}) \label{eq:lkl_BPL} \\
    &\times \mathcal{L}^{\rm (a)}_{\rm IC}(\phi, \gamma_1). \nonumber
\end{align}

We perform a Bayesian joint analysis by defining the posterior $P^{(a)}(\phi, \gamma_1, \gamma_2, \log_{10}E_{\rm b})$ as the product of the likelihood in \cref{eq:lkl_BPL} and the priors on $\phi$, $\gamma_1$, $\gamma_2$, and $\log_{10}E_{\rm b}$. The same uniform priors as in the SPL fit are used for $\phi$ and $\gamma_1$. The prior on $\gamma_2$ is $\pi(\gamma_2; \gamma_1) = \mathcal{U}_{[1, \gamma_1]}(\gamma_2)$, and the prior on the logarithm of the break energy is assumed to be uniform in the considered range: $\pi(\log_{10}E_{\rm b}) = \mathcal{U}_{[6,9]}(\log_{10}E_{\rm b} [\si{\GeV}])$.

We marginalise over $\phi$ and $\gamma_1$ to get the 2D posterior and compute the corresponding contours in the $(\gamma_2, \log_{10}E_{\rm b})$ space, as illustrated in the right panel of \cref{fig:jointfit}. The flux corresponding to the $1\sigma$ contour obtained using the HESE sample is shown in \cref{fig:flux}. The global best-fit value for each parameter, and the corresponding $1\sigma$ uncertainty, are obtained by profiling the three other parameters.

In the frequentist approach, a negative-log-likelihood test statistic is considered. The 2D contours in the $(\gamma_2, E_{\rm b})$ space are obtained following the profiled Feldman-Cousins approach~\cite{NOvA:2022wnj}, treating $\phi$ and $\gamma_1$ as nuisance parameters. The 1D intervals on $\gamma_2$ and $E_{\rm b}$ are computed with a similar procedure, treating also $E_{\rm b}$ or $\gamma_2$ as a nuisance parameter, respectively. The frequentist best-fit BPL flux parameters are in good agreement with the Bayesian ones in \cref{tab:bestfits}. The values reported for $\phi$ and $\gamma_1$ are consistent with IC SPL best fits.

The tensions are computed in the BPL case, giving values of about $2.9\sigma$ in both the PG and PPC tests, as reported in \cref{tab:tensions}.
These reflect the remaining tension between the UHE datasets despite the additional degrees of freedom.

\subsection{Preference for a break}

It is also instructive to quantify the preference in the data for the presence of a break in the spectrum. This is estimated by comparing the SPL and BPL fits presented in the previous sections.

The Bayes factor is commonly used to quantify such a preference. It is defined as the ratio of the evidences $\mathds{E}=\int_{\Theta} P(\Theta) {\rm d}\Theta$ of the two scenarios: $\mathcal{B} = \mathds{E}_{\rm BPL}/\mathds{E}_{\rm SPL}$. The obtained value can be interpreted using Jeffreys' scale~\cite{Jeffreys:1998}, where $\mathcal{B} > 3.2$ ($\mathcal{B} < 0.3$) corresponds to a substantial preference for BPL (SPL). The Bayes factor is computed by integrating the posterior distribution over the parameter space using the nested sampling Monte Carlo algorithm as implemented in the \texttt{UltraNest} package~\cite{Buchner:2021cql}.

Alternatively, the preference for a BPL flux model can be estimated in a frequentist approach by means of a likelihood ratio (LR) test, for which the test statistic is given by:
\begin{align}
    t^{(a)}_{\rm LR} = 2 \log \dfrac{\mathcal{L}^{(a)}_{\rm BPL}(\hat{\Theta}_{\rm BPL})}{\mathcal{L}^{(a)}_{\rm SPL}(\hat{\Theta}_{\rm SPL})},
\end{align}
where $\hat{\Theta}_{\rm SPL}$ and $\hat{\Theta}_{\rm BPL}$ represent respectively the parameters that maximise the likelihoods in \cref{eq:lkl_SPL} and \cref{eq:lkl_BPL}. A p-value is extracted from $t^{(a)}_{\rm LR}$ using Wilks' theorem for two d.o.f., given by the difference in the number of free parameters between the SPL and BPL models. A low p-value indicates a preference for the BPL model. The results from both approaches and for all considered IC samples are reported in \cref{tab:preference}. We find no significant preference for the BPL hypothesis.

\begin{table}[hbtp]
    \caption{Obtained preferences for a broken-power-law spectrum versus the single-power-law hypothesis.\label{tab:preference}}
    \begin{ruledtabular}
    \begin{tabular}{c|ccc|ccc}
        UHE Sample(s) & \multicolumn{3}{c|}{KM3-230213A} & \multicolumn{3}{c}{Global} \\
        \midrule
        HE Sample & HESE & ESTES & NST & HESE & ESTES & NST \\
        \midrule
        Bayes factor $\mathcal{B}$ & 27.0 & 8.7 & 3.9 & 1.2 & 0.6 & 0.3 \\
        LR p-value (\%)& 0.4 & 1.7 & 5.9 & 33 & 86 & 100\\
    \end{tabular}
    \end{ruledtabular}
\end{table}

\section{Joint fit of IC HE and KM3-230213A}
\label{sec:km3netfit}

For completeness, we perform in the following a joint fit of each of the IC HE samples and KM3-230213A. Similarly to what is done in \cref{sec:jointfit}, we consider SPL and BPL flux models spanning from TeV to EeV energies, and evaluate the preference for either of the two models. The likelihood is computed according to \cref{eq:lkl_SPL} and \cref{eq:lkl_BPL} for each flux hypothesis respectively, now defining $\mathcal{L}_{\rm UHE}(\Theta) = \mathcal{L}_{\rm KM3}(\Theta)$.

The corresponding best-fit parameters and tensions are reported in \cref{tab:bestfits} and \cref{tab:tensions}, respectively. Both Bayes factor and LR approaches indicate strong preference for a BPL flux, independently of the IceCube samples considered, as shown in \cref{tab:preference}. This scenario strongly overshoots IC-EHE and Auger constraints in this region, as suggested by the grey cross in \cref{fig:flux}. The BPL best-fit flux would give $80 - 90$ ($100 - 110$) expected events in IC-EHE (Auger) in the analysed energy range, corresponding to a $12\sigma-13\sigma$ ($14\sigma$) tension considering their non-observations.

\section{Conclusion} 
\label{sec:conclusion}

In this article, we study the compatibility of previous analyses carried out by IC and Auger with the detection of KM3-230213A. In this work, a tension in the $2.5\sigma-3\sigma$ range is found. This is in line with the interpretation of the KM3NeT event being a possible upward fluctuation from a flux with a normalisation $E^2 \Phi^{\rm 1f}_{\nu + \bar \nu} = 7.5 \times 10^{-10}~{\rm GeV cm^{-2} s^{-1} sr^{-1}}$, assuming an $E^{-2}$ spectrum in the \SI{72}{\peta\eV} -- \SI{2.6}{\exa\eV} range, as already hinted in Ref.~\cite{Nature}.

Additionally, we consider a joint analysis that includes IC measurements at lower energies. We compare two alternative hypotheses: (a) the KM3-230213A observation is consistent with an extrapolation of the SPL measurements from HESE, ESTES, or NST samples, characterised by a fixed normalisation and a unique spectral index; (b) there is a break in the neutrino-energy distribution, originating from the emergence of a new component, such as cosmogenic neutrinos or a different production process. This second hypothesis is parametrised as a BPL flux, characterised by the same two parameters as in (a) and two additional parameters: the break energy $E_{\rm b}$ and the spectral index after the break $\gamma_2$. The large energy of KM3-230213A pulls the SPL fits towards the IC global best-fit spectral index of about $2.5 - 2.6$. The BPL fit with the HESE constraint shows a best-fit break energy in the \SIrange{1}{10}{\peta\eV} range, with a spectral index of about $2 - 2.3$ above the break, as illustrated in \cref{fig:flux}.

Then, we assess the preference between hypotheses (a) and (b) using both frequentist and Bayesian approaches. Only when using the constraints from the HESE sample, the data shows a slight preference for the presence of a break, although not significant. Overall, the observation of KM3-230213A does not provide enough statistical evidence for the hardening of the diffuse neutrino flux.

Finally, relaxing the UHE constraints set by IC and Auger, we investigate how the flux corresponding to the KM3NeT observation compares to the IC HE measurements. In this context, we assess the preference for either SPL and BPL flux models, and consistently find hints of a break in the diffuse neutrino flux spectrum, with $\gamma_2 \sim 1$ and $E_{\rm b} \sim$ \SIrange{10}{50}{\peta\eV}. However, this is in tension with limits from IC-EHE and Auger; the corresponding BPL best fit suggests a $12-14\sigma$ underfluctuation for the null observations. Further measurements are thus essential to confirm or refute the emergence of a new flux component in the UHE region.

In summary, our global study demonstrates the feasibility and power of combining measurements of various experiments to interpret UHE data in light of observations at lower energies. Our analysis relies on publicly-available data from IC and Auger, leading to some approximations in the treatment of the backgrounds and systematic uncertainties. Furthermore, it is based on simplistic parametrisations of the diffuse neutrino flux and does not account for the possible presence of more complex features. Within these limitations, current data do not yet allow us to precisely constrain the shape of the spectrum, e.g., the presence of a break with respect to well-constrained flux measurements in the TeV--PeV region. In particular, the absence of events in IC and Auger datasets above tens of PeV and the measurement of only one event in KM3NeT leave room for further investigations of the true UHE neutrino spectrum. This study is intended to be a step toward combining multiple measurements to characterise the diffuse astrophysical neutrino spectrum at the highest energies. Future observations, with a larger KM3NeT/ARCA detector configuration~\cite{KM3NeT:2024paj} and with increased exposures from IC and Auger, and upcoming radio instruments exploiting the Askaryan effect to detect UHE neutrinos~\cite{RNO-G:2020rmc,GRAND:2018iaj}, will be crucial to eventually discriminate between different components in the spectrum.

\begin{acknowledgements}
The authors acknowledge the financial support of:
KM3NeT-INFRADEV2 project, funded by the European Union Horizon Europe Research and Innovation Programme under grant agreement No 101079679;
Funds for Scientific Research (FRS-FNRS), Francqui foundation, BAEF foundation.
Czech Science Foundation (GAČR 24-12702S);
Agence Nationale de la Recherche (contract ANR-15-CE31-0020), Centre National de la Recherche Scientifique (CNRS), Commission Europ\'eenne (FEDER fund and Marie Curie Program), LabEx UnivEarthS (ANR-10-LABX-0023 and ANR-18-IDEX-0001), Paris \^Ile-de-France Region, Normandy Region (Alpha, Blue-waves and Neptune), France,
The Provence-Alpes-Côte d'Azur Delegation for Research and Innovation (DRARI), the Provence-Alpes-Côte d'Azur region, the Bouches-du-Rhône Departmental Council, the Metropolis of Aix-Marseille Provence and the City of Marseille through the CPER 2021-2027 NEUMED project,
The CNRS Institut National de Physique Nucléaire et de Physique des Particules (IN2P3);
Shota Rustaveli National Science Foundation of Georgia (SRNSFG, FR-22-13708), Georgia;
This work is part of the MuSES project which has received funding from the European Research Council (ERC) under the European Union’s Horizon 2020 Research and Innovation Programme (grant agreement No 101142396);
The General Secretariat of Research and Innovation (GSRI), Greece;
Istituto Nazionale di Fisica Nucleare (INFN) and Ministero dell’Universit{\`a} e della Ricerca (MUR), through PRIN 2022 program (Grant PANTHEON 2022E2J4RK, Next Generation EU) and PON R\&I program (Avviso n. 424 del 28 febbraio 2018, Progetto PACK-PIR01 00021), Italy; IDMAR project Po-Fesr Sicilian Region az. 1.5.1; A. De Benedittis, W. Idrissi Ibnsalih, M. Bendahman, A. Nayerhoda, G. Papalashvili, I. C. Rea, A. Simonelli have been supported by the Italian Ministero dell'Universit{\`a} e della Ricerca (MUR), Progetto CIR01 00021 (Avviso n. 2595 del 24 dicembre 2019); KM3NeT4RR MUR Project National Recovery and Resilience Plan (NRRP), Mission 4 Component 2 Investment 3.1, funded by the European Union – NextGenerationEU, CUP I57G21000040001, Concession Decree MUR No. n. Prot. 123 del 21/06/2022;
Ministry of Higher Education, Scientific Research and Innovation, Morocco, and the Arab Fund for Economic and Social Development, Kuwait;
Nederlandse organisatie voor Wetenschappelijk Onderzoek (NWO), the Netherlands;
The grant “AstroCeNT: Particle Astrophysics Science and Technology Centre”, carried out within the International Research Agendas programme of the Foundation for Polish Science financed by the European Union under the European Regional Development Fund; The program: “Excellence initiative-research university” for the AGH University in Krakow; The ARTIQ project: UMO-2021/01/2/ST6/00004 and ARTIQ/0004/2021;
Ministry of Research, Innovation and Digitalisation, Romania;
Slovak Research and Development Agency under Contract No. APVV-22-0413; Ministry of Education, Research, Development and Youth of the Slovak Republic;
MCIN for PID2021-124591NB-C41, -C42, -C43 and PDC2023-145913-I00 funded by MCIN/AEI/10.13039/501100011033 and by “ERDF A way of making Europe”, for ASFAE/2022/014 and ASFAE/2022 /023 with funding from the EU NextGenerationEU (PRTR-C17.I01) and Generalitat Valenciana, for Grant AST22\_6.2 with funding from Consejer\'{\i}a de Universidad, Investigaci\'on e Innovaci\'on and Gobierno de Espa\~na and European Union - NextGenerationEU, for CSIC-INFRA23013 and for CNS2023-144099, Generalitat Valenciana for CIDEGENT/2018/034, /2019/043, /2020/049, /2021/23, for CIDEIG/2023/20, for CIPROM/2023/51 and for GRISOLIAP/2021/192 and EU for MSC/101025085, Spain;
Khalifa University internal grants (ESIG-2023-008, RIG-2023-070 and RIG-2024-047), United Arab Emirates;
The European Union's Horizon 2020 Research and Innovation Programme (ChETEC-INFRA - Project no. 101008324);
C. A. Arg\"uelles and N. W. Kamp were supported by the David \& Lucille Packard Foundation; A. Wen was supported by the Natural Sciences and Engineering Research Council of Canada (NSERC), funding reference number PGSD-577971-2023.
\end{acknowledgements}

\section*{Author contributions}

The KM3NeT research infrastructure is being built, operated and maintained by the KM3NeT Collaboration. Authors contributed to the design, construction, deployment and operation of the detector, as well as to the data taking, detector calibration, development of the software, data processing, Monte Carlo simulations and various analyses required for this work. The main authors of this analysis are: C.~Arg\"uelles, G.~de Wasseige, N.~Kamp, M.~Lamoureux, J.~Mauro, P.~Sevle Myhr, and A.~Y.~Wen. The final manuscript was reviewed and approved by all authors.

\bibliography{references}

\clearpage
\appendix

\begin{center}
\textbf{\large Appendices}
\end{center}

\section{Statistical methods and notation}
\label{sec:stat}

In this article, the per-flavour $\nu+\bar\nu$ astrophysical flux is assumed to be isotropic and steady, such that it is uniquely defined by its energy distribution: ${\rm d}^3N/({\rm d}E {\rm d}t {\rm d}\Omega) = \Phi^{\rm 1f}_{\nu+\bar\nu}(E; \Theta)$ in units of \si{\per\GeV\per\square\cm\per\second\per\steradian}, where $\Theta$ denotes the various flux parameters.

To characterise the UHE observations, two independent energy ranges are used throughout the analysis. The central 90\% energy range of KM3-230213A, $\mathcal{I}_{\rm in}(\Theta)$, is defined by estimating the 5th and 95th percentile of the neutrino energy distribution given the measured muon energy and the spectral shape parametrised by $\Theta$. The complementary interval is $\mathcal{I}_{\rm out}^{d}(\Theta) = \left[E_{\rm min}^{d}, E_{\rm max}^{d}\right] \setminus \mathcal{I}_{\rm in}(\Theta)$, where $E_{\rm min}^{d}$ and $E_{\rm max}^{d}$ are extremes specific to each UHE dataset. For KM3NeT, $E_{\rm min}^{\rm KM3} = \SI{100}{\tera\eV}$, $E_{\rm max}^{\rm KM3} = \SI{100}{\exa\eV}$.

Therefore, the number of expected events associated with a given UHE dataset in an energy range $\mathcal{I}_E$ is given by
\begin{equation}
    \mu_{\rm exp}^{d}(\Theta; \mathcal{I}_E) = \int_{{\mathcal{I}_E}} \mathcal{E}^{d}(E) \times \Phi^{\rm 1f}_{\nu+\bar\nu}(E; \Theta) \, {\rm d}E,
    \label{eq:nexpected}
\end{equation}
where $\mathcal{E}^{d}(E)$ is the experiment-dependent exposure. Furthermore, for the UHE analyses considered here, the contribution from background events is neglected.

The analyses presented in this article rely on a simple Poisson counting to convert UHE observations into constraints on the neutrino flux parameters $\Theta$:
\begin{align}
    \mathcal{L}_{d,\rm in}(\Theta) &\equiv \mathrm{Poisson} \left(N_{\rm obs}^{d,\rm in}; \mu_{\rm exp}^{d}(\Theta; \mathcal{I}_{\rm in}(\Theta))\right)\label{eq:lkl_in} \\
    \mathcal{L}_{d,\rm out}(\Theta) &\equiv \mathrm{Poisson} \left(N_{\rm obs}^{d,\rm out}; \mu_{\rm exp}^{d}(\Theta; \mathcal{I}^d_{\rm out}(\Theta))\right)\label{eq:lkl_out}\\
    \mathcal{L}_{d}(\Theta) &\equiv \mathcal{L}_{d,\rm in}(\Theta) \times \mathcal{L}_{d,\rm out}(\Theta) \label{eq:lkl}
\end{align}
where $\mathrm{Poisson}(N; \lambda) = e^{-\lambda} \lambda^{N}/N!$, $d$ denotes any of the UHE datasets (KM3NeT, IC-EHE, Auger), and $N_{\rm obs}^{d,\rm in}$ and $N_{\rm obs}^{d,\rm out}$ represent the number of observed events in the intervals $\mathcal{I}_{\rm in}$ and $\mathcal{I}^d_{\rm out}$ respectively. In the case of KM3NeT, we have $N_{\rm obs}^{\rm KM3,in} = 1$, and $N_{\rm obs}^{\rm KM3,out} = 0$.

\section{External datasets} 
\label{sec:datasets}

\paragraph*{IC-EHE}--- The IC Extremely-High-Energy exposure $\mathcal{E}^{\rm IC-EHE}$ is extracted from the 9-year analysis described in \cite{IceCube:2018fhm}, for which the effective area is shown in \cite{Meier:2024flg} (from \SI{1}{\peta\eV} to \SI{50}{\exa\eV}). The analysis reports only neutrino candidates around \SI{10}{\peta\eV}. As the handling of these events would require careful treatment of the expected background and systematics, the integration range is chosen to start at higher energies, with $E_{\rm min}^{\rm IC-EHE} = \SI{20}{\peta\eV}$, $E_{\rm max}^{\rm IC-EHE} = \SI{50}{\exa\eV}$, and $N_{\rm obs}^{\rm IC-EHE,in} = N_{\rm obs}^{\rm IC-EHE,out} = 0$.

\paragraph*{Auger}--- The Auger sky-averaged exposure $\mathcal{E}^{\rm Auger}$ is computed considering their three samples (Earth-skimming, low-zenith downward-going, high-zenith downward-going), using the effective area from the data release associated to \cite{PierreAuger:2019azx} and the 18-yr livetime reported in the most recent analysis~\cite{PierreAuger:2023pjg}, with $E_{\rm min}^{\rm Auger} = \SI{10}{\peta\eV}$, $E_{\rm max}^{\rm Auger} = \SI{1}{\zetta\eV}$. No neutrino events have been identified in any of the samples: $N_{\rm obs}^{\rm Auger,in} = N_{\rm obs}^{\rm Auger,out} = 0$.

\paragraph*{IC high-energy measurements}--- The measurements performed below tens of PeV with selected IC samples are also considered: High-Energy Starting Events (HESE, \cite{IceCube:2020wum}), Enhanced Starting Track Event Selection (ESTES, \cite{IceCube:2024fxo}), and Northern-Sky Tracks (NST, \cite{IceCube:2021uhz}). For simplicity, the single-power-law measurements ($\phi \times (E/\SI{100}{\tera\eV})^{-\gamma_1}$) from these three samples are extracted from the scans reported in the related publications:
\begin{equation}
    \textrm{TS}^{(a)}_{\rm IC}(\phi, \gamma_1) = -2\log \mathcal{L}^{(a)}_{\rm IC}(\phi, \gamma_1).
    \label{eq:ts_ic}
\end{equation}
The scans are typically defined in the ranges $\phi \in [1, 4] \times \SI{e-18}{\per\GeV\per\square\cm\per\second\per\steradian}$ and $\gamma \in [1.5, 3.5]$, motivating the choices of priors in the Bayesian fits.

\begin{figure}[hbtp]
    \centering
    \includegraphics[width=\linewidth]{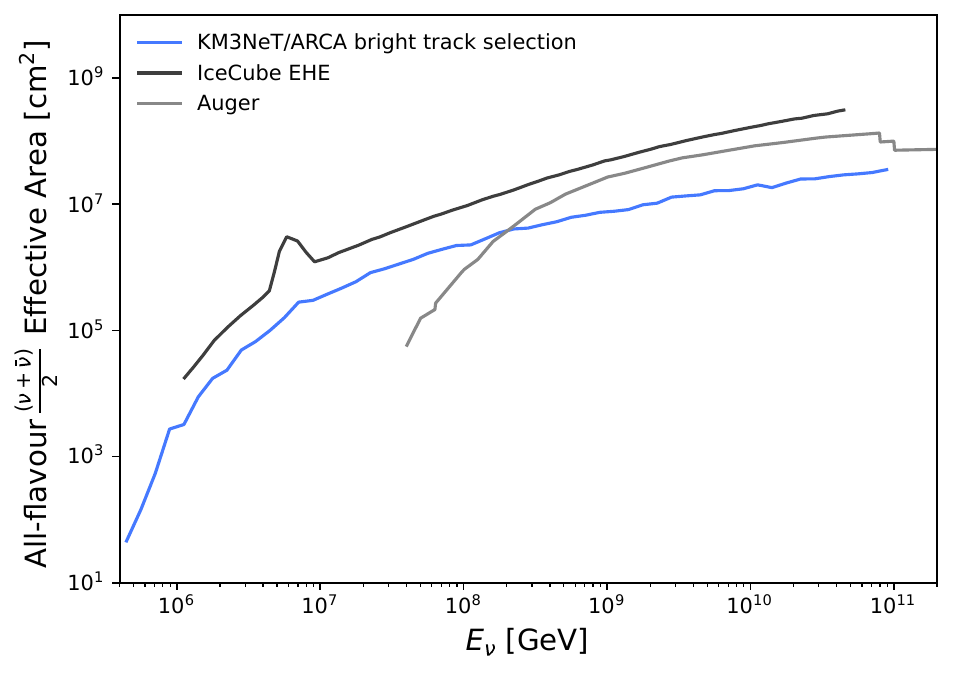}
    \caption{Sky-averaged all-flavour effective areas of KM3NeT/ARCA bright track selection~\cite{Nature}, IC-EHE~\cite{Meier:2024flg}, and Auger~\cite{PierreAuger:2019azx} (summed over Earth-skimming, low-zenith downward-going, high-zenith downward-going samples), averaged between neutrinos and antineutrinos.\label{fig:aeff}}
\end{figure}

The effective areas of the three UHE samples are illustrated in \cref{fig:aeff}.

\end{document}